\documentclass[twocolumn,showpacs,aps,prl,showkeys,superscriptaddress]{revtex4-1}

\usepackage{amssymb,amsmath,amsfonts}
\usepackage{graphicx}
\usepackage{xcolor}
\usepackage{xspace}
\usepackage{ulem}
\usepackage[colorlinks=true, urlcolor=blue]{hyperref}
\usepackage{wasysym}
\usepackage{siunitx}
\usepackage{rotating}

\def\gsim{\raise0.3ex\hbox{$\;>$\kern-0.75em\raise-1.1ex\hbox{$\sim\;$}}}
\def\lsim{\raise0.3ex\hbox{$\;<$\kern-0.75em\raise-1.1ex\hbox{$\sim\;$}}}

\newcommand{\ba}[1]{\begin{eqnarray} \label{(#1)}}
\newcommand{\ea}{\end{eqnarray}}

\newcommand{\AddrUTFSM}{
Departmento de F\'isica y CCTVal\\
Universidad T\'ecnica Federico Santa Mar\'ia, Valpara\'iso 2340000, Chile}

\newcommand{\AddrUNAB}{
Departamento de Ciencias F\'isicas, Universidad Andres Bello, \\
Sazie 2212, Piso 7, Santiago, Chile.}

\newcommand{\AddrULS}{
Departamento de F\' isica y Astronom\' ia, Facultad de Ciencias, Universidad de La Serena, \\
Avenida Cisternas 1200, La Serena, Chile.  
}
 
\newcommand{\AddrTLV}{
School of Physics and Astronomy, Tel Aviv University, \\ Tel Aviv 69978, Israel.
}

\def\gsim{\raise0.3ex\hbox{$\;>$\kern-0.75em\raise-1.1ex\hbox{$\sim\;$}}}
\def\lsim{\raise0.3ex\hbox{$\;<$\kern-0.75em\raise-1.1ex\hbox{$\sim\;$}}}

\begin{document}

\title{Searching for a Sterile Neutrino in Tau Decays at B-factories}

\author{C.O.\, Dib}\email{claudio.dib@usm.cl}\affiliation{\AddrUTFSM}
\author{J.C. Helo} \email{jchelo@userena.cl}\affiliation{\AddrULS}
\author{M. Nayak}
\email{minakshi@tauex.tau.ac.il}\affiliation{\AddrTLV}
\author{N.A.\,  Neill}\email{nicolas.neill@gmail.com}\affiliation{\AddrUTFSM}
\author{A. Soffer}
\email{asoffer@tau.ac.il}\affiliation{\AddrTLV}
\author{J. Zamora-Saa} \email{jilberto.zamora@unab.cl}\affiliation{\AddrUNAB}

\keywords{neutrino masses and mixing; tau decays}
\pacs{14.60.Pq,  12.60.Jv,  14.80.Cp}

\begin{abstract}
The phenomenon of neutrino flavor oscillations motivates searches for sterile neutrinos in a broad range of masses and mixing-parameter values. A sterile neutrino $N$ that mixes predominantly with the $\tau$ neutrino is particularly challenging  experimentally. To address this challenge, we propose a new method to search for a $\nu_\tau$-mixing  with $N$  lighter than the $\tau$ lepton. The method uses the large $e^+e^-\to\tau^+\tau^-$ samples collected at $B$-factory experiments to produce the $N$ in $\tau$-lepton decays. We exploit the long lifetime of a sterile neutrino in this mass range to suppress background and apply kinematic and vertexing constraints that enable measuring the sterile neutrino mass. Estimates for the sensitivities of the BaBar, Belle, and Belle~II experiments are calculated  and presented.
\end{abstract}

\maketitle

The discovery of neutrino flavor oscillations~\cite{Fukuda:1998mi, Ahmad:2002jz, Eguchi:2002dm} has opened a new frontier in the search for new physics. Oscillations necessitate that at least two neutrinos have mass, while in the standard model (SM) of electroweak interactions neutrinos are massless~\cite{Tanabashi:2018oca}. Neutrino masses can be incorporated into the SM framework via Yukawa interactions, the mechanism that gives mass to all other fundamental fermions, by introduction of right-handed (RH) neutrino fields. Being neutral under the SM interactions, the RH fields can have Majorana mass terms, which lead to new observable phenomena. These include breaking of lepton-number symmetry, and the existence of additional neutrino mass eigenstates, with masses anywhere from a few eV all the way to grand unification scales, $O(\SI{e15}{\GeV})$. These extra mass eigenstates, composed mostly of the RH states, are referred to as ``sterile" neutrinos.

Various experiments have set bounds on specific combinations of masses and mixing parameters of the sterile neutrinos.
Most recently, the ATLAS and CMS collaborations at the LHC have searched for sterile neutrinos in final states with like-sign dileptons, two jets and no missing energy  \cite{Khachatryan:2016olu, Aad:2015xaa}, as well as  three leptons~\cite{Sirunyan:2018mtv, Aad:2019kiz}. These searches put bounds on mixing parameters for sterile-neutrino masses $m_N$ above a few GeV and up to about 1~TeV. 
Searches for sterile neutrinos with masses of the order of GeV have also been proposed for various meson decays~\cite{Zhang:2010um, Yuan:2013yba, Helo:2010cw, Dib:2000wm, Cvetic:2010rw, Dib:2011jh, Cvetic:2012hd, Bonivento:2013jag, Cvetic:2013eza, Dib:2014pga, Lees:2013gdj, Dong:2013raa,Cvetic:2014nla,Cvetic:2015ura,Zamora-Saa:2016qlk}, tau lepton decays~\cite{Gribanov:2001vv, Castro:2012gi,Zamora-Saa:2016ito, Helo:2011yg, Dib:2011hc}, top quarks decays~\cite{Quintero:2011yh} and W boson decays~\cite{Helo:2013esa, Dib:2015oka, Dib:2016wge,Cvetic:2018elt, Antusch:2017hhu,Cottin:2018nms,Abada:2018sfh}.

Here we study the case in which an on-shell sterile neutrino $N$ is produced in the $\tau$-lepton decays $\tau\to N X_1$, where $X_1$ describes all possible SM final states allowed by kinematics and conserved quantum numbers. This search is uniquely sensitive to an 
$N$ that mixes predominantly with the $\tau$ neutrino and is lighter than the $\tau$ lepton, i.e., $m_N<m_\tau=1.777$~GeV. The best existing limits for this  case arise from a lepton-flavor-agnostic search performed by the DELPHI experiment~\cite{Abreu:1996pa}. We show that tighter limits can be obtained with the data sets collected by the $B$-factory experiments BaBar~\cite{Aubert:2001tu} and Belle~\cite{Abashian:2000cg} at center-of-mass (CM) energies $\sqrt{s}=10.58$~GeV. BaBar and Belle have collected, respectively, ${\cal N}_{\tau \tau}= 4.6\times 10^{8}$ and $8.8\times 10^{8}$ easily identifiable $e^+e^-\to\tau^+\tau^-$ events. The Belle~II experiment~\cite{Abe:2010gxa}, which is scheduled to collect ${\cal N}_{\tau \tau}=4.6\times 10^{10}$ $e^+e^-\to\tau^+\tau^-$ events by about 2027, will achieve even greater sensitivity.

Our method exploits the long lifetime of the low-mass $N$, which goes as $\tau_N\sim m_N^{-5}$. Particularly when produced at relativistic speeds, the $N$ travels a macroscopic distance inside the detector before  decaying~\cite{Helo:2013esa, Dib:2014iga, Shrock:1978ft}. The resulting displaced-vertex signature is particularly useful for suppressing backgrounds. This has been exploited for a variety of new-particle searches~\cite{Lee:2018pag,Alimena:2019zri}, but so far only twice at $B$~factories~\cite{Liventsev:2013zz,Lees:2015rxq}. Our approach is different from that of Refs.~\cite{Gribanov:2001vv, Castro:2012gi,Zamora-Saa:2016ito, Helo:2011yg, Dib:2011hc}, which rely primarily on violation of lepton-number, lepton-flavor, or CP symmetries, and from that of Ref.~\cite{Kobach:2014hea}, which  uses the kinematics of $\tau$ decays to search for the $N$ and is not sensitive for $m_N$ larger than a kinematic endpoint. Various proposed, dedicated long-lived-particle experiments would be sensitive to a long-lived $N$~\cite{Alekhin:2015byh, Gligorov:2017nwh, Kling:2018wct, Helo:2018qej, Curtin:2018mvb, Dercks:2018wum, SHiP:2018xqw,Gligorov:2018vkc}.
However, in most cases they lack the ability to identify the dominant $\nu_\tau$ mixing, particularly for the case $m_N<m_\tau$.

As theoretical framework, we use the generic form of seesaw models \cite{Minkowski:1977sc,Yanagida:1979as,GellMann:1980vs,Mohapatra:1980yp,Glashow:1979nm,Schechter:1980gr, Mohapatra:1986bd}, where the SM neutrinos
$\nu_\ell$ ($\ell= e,\mu, \tau$) are mainly the light fields $\nu_i$, with small admixtures of 
extra fields $N_j$, which are heavier and sterile under the SM gauge interactions:
\begin{equation}
\nu_{\ell} = \sum_{i=1}^3 U_{\ell i} \nu_{i} + \sum_j V_{\ell N_j} N_j.
\label{admixture}
\end{equation}
While each seesaw model contains specific relations between masses and mixings, here we take a more model-independent approach where 
$m_{N_j}$ and $V_{\ell N_j}$ are independent parameters. We focus on scenarios where one of the sterile neutrinos $N_j$ can be produced in $\tau$ lepton decays, i.e. $m_{N_j} < m_\tau$.
Henceforth we discard the index $j$ to refer to that neutrino. Moreover, we study the case $|V_{\tau N}|\gg |V_{e N}|,\,   |V_{\mu N}|$, in which the $N$ mixes mainly with the $\tau$ neutrino, and its mixing with the electron or muon neutrinos can be neglected. 
Particular interest in this scenario stems from the fact that 
existing limits on $|V_{eN}|$ and $|V_{\mu N}|$ are much tighter than those on $|V_{\tau N}|$, 
because electrons and muons are  experimentally easier to identify than $\tau$ leptons.
Thus, our simplified model is described by the effective Lagrangian
\begin{align}
\mathcal{L}  = &
  - \frac{g}{\sqrt{2}} W_\mu^+ V^*_{\tau N} \overline{N} \gamma^\mu P_L \tau +\mbox{h.c.}\nonumber\\
& - \frac{g}{2\cos\theta_W} Z_\mu V^*_{\tau N} \overline{N} \gamma^\mu P_L \nu_\tau +\mbox{h.c.},
\label{eq:lag}
\end{align}
where $g$ is the SM electroweak coupling constant, $\theta_W$ is the 
Weinberg angle, $W^+_\mu$ and $Z_\mu$ are the heavy electroweak gauge boson fields, and $P_L$ is the left-handed projection operator.
\begin{figure}[t]
\begin{minipage}[b]{.7\linewidth}
\includegraphics[width=\linewidth]{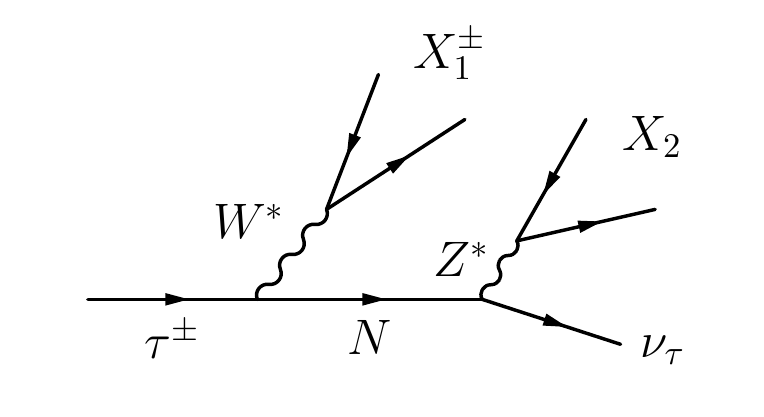}
\end{minipage}
\caption{The proposed decay chain, $\tau\to X_1 N$ followed by $N\to \nu_\tau X_2$.}
\label{fig:fd} 
\end{figure}

To probe this scenario, we propose to search for a long-lived $N$ produced via $\tau^-\to X_1^- N$, taking advantage of copious $e^+  e^-\to \tau^+\tau^-$ events at $B$-factory experiments.  
Since we study the case with dominant $V_{\tau N}$ mixing, the only sizeable charged-current decay of the $N$ is $N\to \tau W^*$.
However, this decay is kinematically forbidden by the condition $m_N<m_\tau$. Therefore, the $N$ must decay via the neutral-current decay  $N\to \nu_\tau X_2$, mediated by the $Z_\mu$ term of Eq.~(\ref{eq:lag}). The complete decay chain is shown in Fig.~(\ref{fig:fd}). 

Our aim here is to estimate the sensitivity of the proposed $B$-factory search to the mixing $|V_{\tau N}|^2$. The number of observed events should be given by:
\begin{equation}\label{nevents}
{\cal N} = {\cal N}_{\tau \tau}\times {\cal B}(\tau\rightarrow X_1 N)\times {\cal B}(N\rightarrow \nu_\tau X_2) \times a \times \epsilon,
\end{equation}
\normalsize
where ${\cal N}_{\tau \tau}$ is the total number of tau lepton pairs produced, ${\cal B}$ denotes a branching fraction, $a$ is the acceptance (which is essentially the probability for $N$ to decay inside the detector), and $\epsilon$ is the reconstruction efficiency. 
We focus on the case in which $X^\pm_1$ is either $\pi^\pm$ or $\pi^\pm \pi^0$. Limiting $X_1$ to hadronic states facilitates the application of the constraints discussed below, which greatly reduce the backgrounds. Use of the leptonic modes $X_1=\ell\nu$ roughly doubles the exploited $\tau$ branching fraction, and is recommended for the actual data analysis. Since these modes do not satisfy the above-mentioned constraints, their study requires full detector simulation, which is beyond the scope of the current study.  
In addition, three-pion and four-pion final states may be used to further increase the sensitivity. The branching fractions ${\cal B}(\tau\rightarrow \pi N)$ and ${\cal B}(\tau\rightarrow \pi \pi N)$ can be obtained after replacing $N\rightarrow \tau$ and $\ell \rightarrow N$ in Eqs.~(3) and (5) of Ref.~\cite{1801.03624}.

We further focus our study on the case in which $X_2$ is $e^+e^-$ or $\mu^+\mu^-$, and use the corresponding branching fractions ${\cal B}(N\rightarrow \nu_\tau X_2)$, obtained from Ref. \cite{1805.08567}. These leptonic branching fractions are approximately 2\% each, for $m_N\gtrsim 800$ MeV.
Hadronic $X_2$ final states with at least two charged pions (needed in order to clearly detect a displaced vertex) are recommended for the actual data analysis, and we comment on them below.

To estimate the acceptance $a$ for the Belle~II detector, we generate signal events using EvtGen~\cite{Lange:2001uf} with beam energies $E_{e^-}=7$~GeV, $E_{e^+}=4$~GeV. Events are produced for $N$ mass values $m_N=0.3$, $0.5$, $0.7$, $1.0$, $1.3$, and $1.6$~GeV, with various values of $|V_{\tau N}|^2$, using the $N$ width given in Ref.~\cite{1805.08567}. 
The acceptance is the fraction of events in which the  $N$ decays inside the acceptance volume,  which we define to be $10<r<80$~cm in the transverse plane, $-40<z<120$~cm in the longitudinal direction, and $|\vec r|>10$~cm. The cut $r>10$~cm rejects prompt tracks and most background from $K_S$ decays and particle interactions in dense detector material. The remaining criteria are chosen so that the $N$ decay occurs at least 40~cm from the edges of the Belle~II drift chamber~\cite{Abe:2010gxa}, to allow sufficient distance for accurate tracking. 

We estimate the efficiency and the background level after application of several selection criteria in a future analysis. 
First, events will be required to have only 4 tracks, with relatively little energy in calorimeter clusters that are not associated to the tracks or to $\pi^0$ mesons that are part of the reconstructed $\tau$ candidates. This will efficiently reject $e^+e^-\to B\bar B$ and $e^+e^-\to q\bar q$ events 
($q=u,d,s,c$)
which have high track multiplicity~\cite{Brandenburg:1999cs} and visible energy, leaving  $e^+e^-\to\tau^+\tau^-$ as the dominant background. 
While the track reconstruction efficiency is greater than 90\%~\cite{Allmendinger:2012ch}, we assume an efficiency of only 25\% for finding the 4 tracks, to conservatively account for reduced efficiency  for tracks that originate far from the interaction point  (IP).

Non-$\tau\tau$ events can be suppressed  further, by exploiting the back-to-back production of the $\tau$ leptons in $e^+e^-\to \tau^+\tau^-$ events, with only little impact on signal efficiency. For example, we find  that requiring the cosine of the CM-frame angle between the momentum vector of the $X_1 X_2$ system and that of the decay products of the other $\tau$ in the event to be less than $-0.5$, and requiring the CM energy of the $X_1 X_2$ system to be greater than 3~GeV yields a signal efficiency of about 90\%.

We take the efficiency associated with particle-identification criteria applied to all 4 tracks to be 60\%~\cite{TheBaBar:2013jta}. 
Beyond particle-identification requirements, vertices arising from the decay $K_S\to\pi^+\pi^-$ or from photon conversion can be very efficiently rejected with requirements on the dilepton-candidate invariant mass and on the angle between the dilepton momentum and the vector pointing from the IP to the vertex location, and by vetoing vertices that are inside dense material~\cite{Lees:2015rxq}. We take the efficiency for these requirements to be 90\%. 
Thus, the total signal efficiency is approximately $\epsilon=10\%$ for the case $X_1=\pi^-$. This estimate is valid when the average flight distance is significantly larger than the linear size of the acceptance volume.
We take the efficiency for the $X_1=\pi^\pm\pi^0$ mode to be half of this, to conservatively account for the $\pi^0\to \gamma\gamma$ reconstruction.

After the above requirements, the dominant background is expected to arise from $e^+e^-\to\tau^+\tau^-$ events containing the decay chain $\tau\to \pi K_L \nu_\tau$, $K_L\to \pi\mu\nu$, with the displaced  pion misidentified as a muon. We refer to this as the $K_L$  background. The product branching fraction for this decay chain is 
$1.35\times 10^{-3}$. Generating such events with EvtGen, we find their acceptance to be 1.4\%. One expects them to have an efficiency similar to that of signal, except that pions have an efficiency of order 0.5\% for passing the lepton-identification criteria~\cite{Liventsev:2013zz}. This value can change by a factor of 2 depending on the cuts applied and the performance of the detector. Ignoring such detail, we thus find $a\times\epsilon\approx 7.5 \times 10^{-5}$. Thus, the expected yield of this background is about 17 events per ${\rm ab}^{-1}$, or 850 in the entire Belle~II event sample.

Further background suppression can be obtained by exploiting  the constraints of the signal hypothesis. The decay chain $\tau\to X_1 N$, $N\to X_2 \nu_\tau $ cannot be fully reconstructed, due to the unobservable neutrino in the final state.
As a result, there are 12 unknowns, namely, the 4-momenta $p^\mu _\nu$, $p^\mu_N$, and $p^\mu_\tau$ of the unreconstructed $\nu_\tau$, $N$, and $\tau$,  respectively. However, the decay chain has 12 constraints: 4-momentum conservation in the $\tau$ and $N$ decays (8 constraints), the known masses of the $\tau$ and the $\nu_\tau$ (2 constraints), and the unit vector from the production point of the $X_1$ system to that of the $X_2$ system, which is the direction of $\vec p_N$ (2 constraints). Solving the constraint equations, one determines the 4-momenta of all the particles up to a two-fold ambiguity arising from a quadratic equation. We label the two solutions for the $N$ mass as $m_1$ and $m_2$, and those for the $\tau$ energies in the collider CM frame as $E_1$ and $E_2$. 

\begin{figure}[t]
\begin{tabular}{cc}
\includegraphics[width=0.5\linewidth]{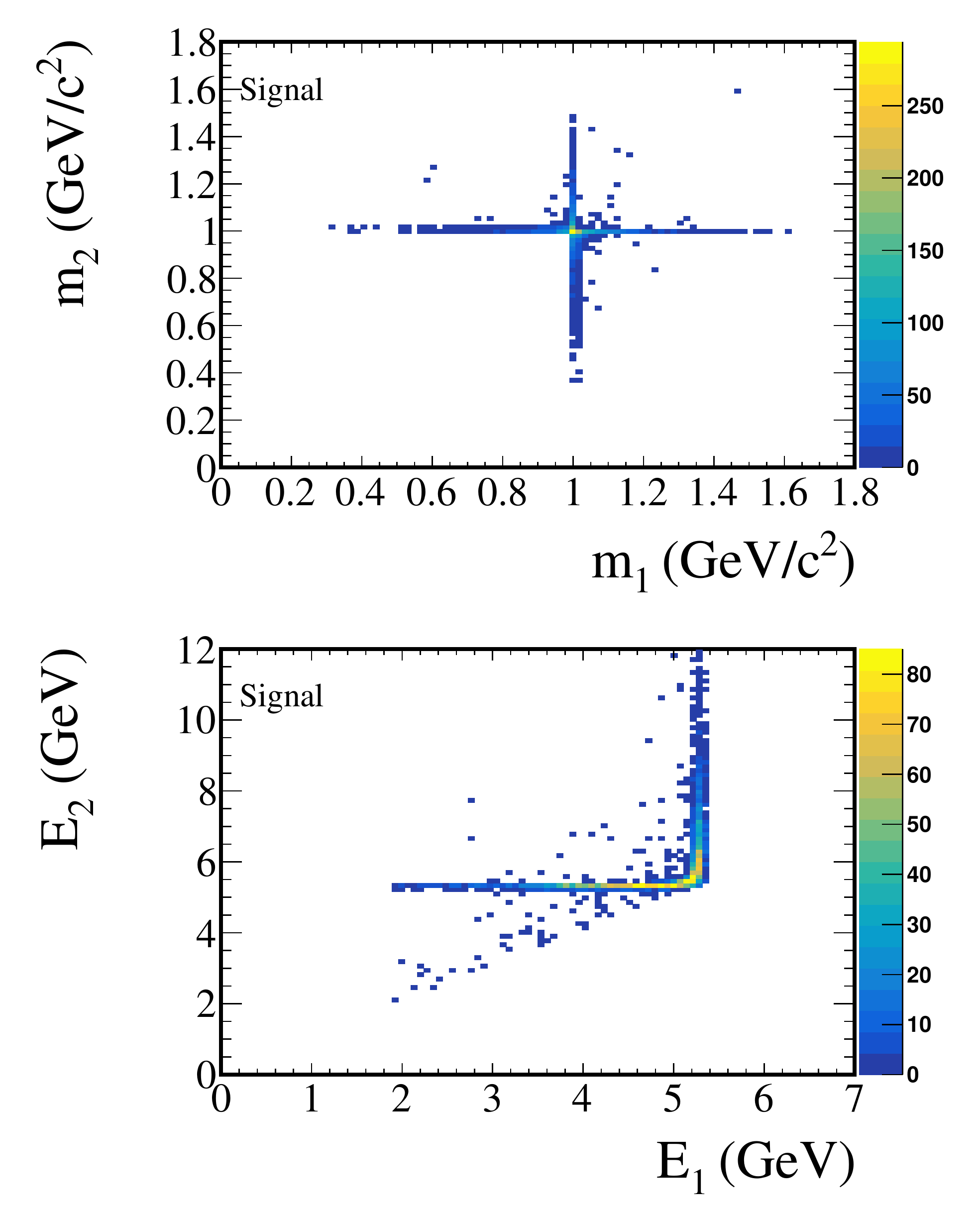}
&
\includegraphics[width=0.5\linewidth]{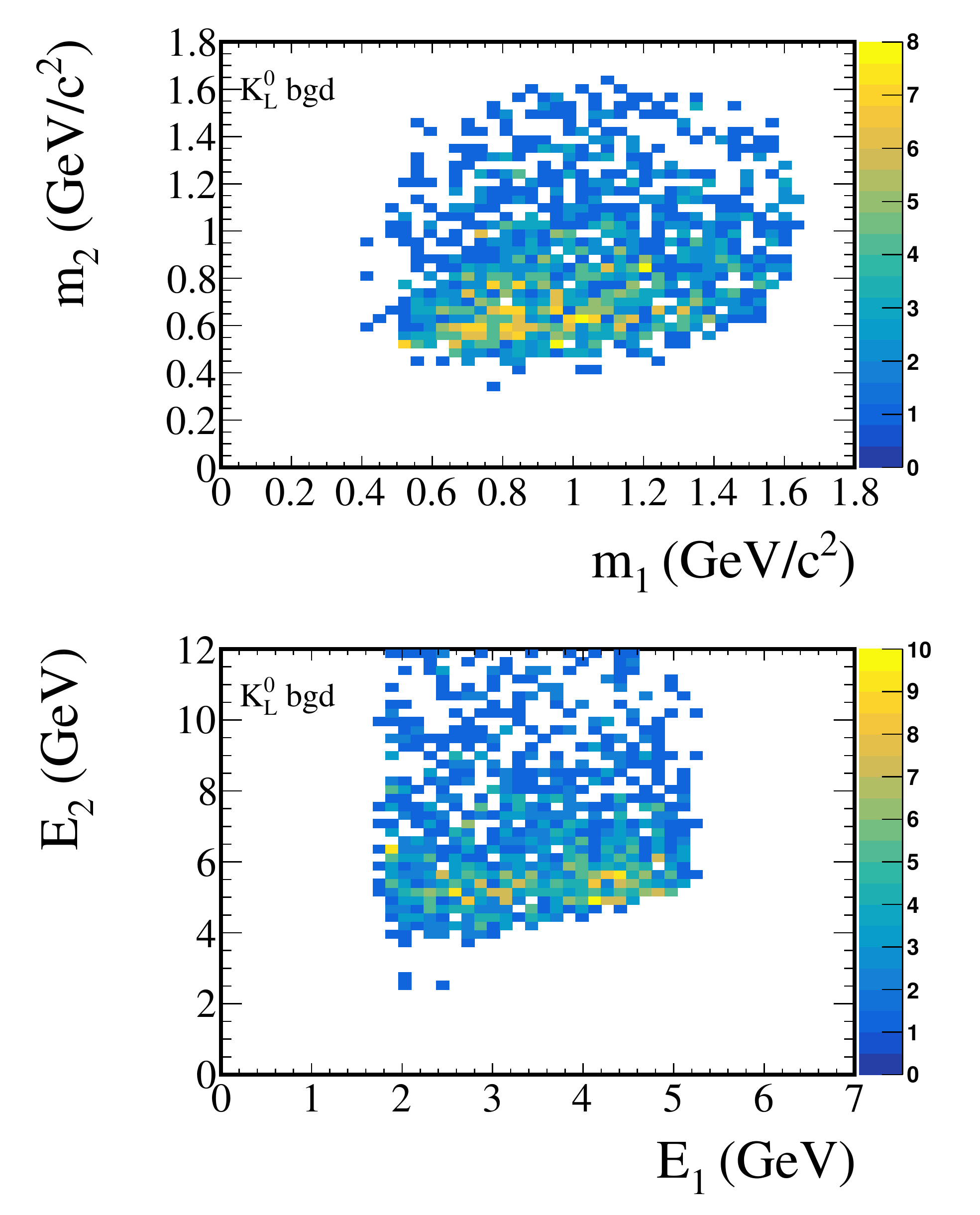}
\end{tabular}
\caption{Distributions of the $N$ mass solutions $m_2$ vs. $m_1$ (top plots) and of the $\tau$ CM-energy solutions $E_2$ vs. $E_1$ (bottom plots) for signal events generated with $m_N=1$~GeV (left plots) and for 
$K_L$ background events (right  plots). Bin color indicates the bin content in arbitrary units.}
\label{fig:distr}  
\end{figure}

The distributions of $m_i$ and $E_i$ for simulated signal and $K_L$ events are shown in Fig.~\ref{fig:distr}. For signal, either $m_1$ or $m_2$ equals the true value of $m_N$, up to some smearing that is due to final-state radiation of photons. Some additional smearing is expected due to detector resolution, not included in this simulation. Similarly, either $E_1$ or $E_2$ peak at the true $\tau$ energy $E_\tau^{\rm true}=\sqrt{s}/2=5.29$~GeV. The $K_L$ background distributions are much broader, as expected due to the additional neutrino. 

The narrow $E_1$ ($E_2$) distribution for signal should be broadened toward lower (higher) values by initial-state radiaion (ISR), which is not simulated in EvtGen. To estimate this effect, we study $e^+ e^-\to \tau^+\tau^-$ + ISR photons generated with KKMC~\cite{Jadach:1999vf}. The $E_\tau^{\rm true}$ distribution of these events has a peak centered at 5.27~GeV with a width of 47~MeV, and a tail due to rare emission of hard ISR photons. After convolving the $E_2$-vs.-$E_1$ distribution with the $E_\tau^{\rm true}$ distribution, we find that requiring  either $E_1>5$~GeV or $5<E_2<5.8$~GeV retains 75\% of the signal while rejecting 75\% of the background. This gives a sense of the background suppression capability provided by $E_1$ and $E_2$.

An optimal way to determine the signal yield in the final data analysis is to fit the $m_2$-vs.-$m_1$ distribution of the data to the sum of a signal distribution plus a background distribution. The signal distribution will be obtained from simulated signal events generated with particular values of $m_N$ and $|V_{\tau N}|^2$. The background distribution can also be obtained from simulation of generic events, validated with control samples. These should be mostly $e^+e^-\to \tau^+\tau^-$ events reconstructed in specific decays, including $\tau\to \pi K_L\nu$ with $K_L\to \pi\mu\nu$ and $\tau\to \pi  K_S\nu$ with $K_S\to \pi^+\pi^-$.
For each point in ($m_N$, $|V_{\tau N}|^2$) space, the fit gives the signal yield and the local signal significance. Various methods exist for converting this into the global significance, which accounts for the a-priori lack of knowledge of $m_N$~\cite{Gross:2010qma}.

\begin{figure}[t]
\includegraphics[width=\linewidth]{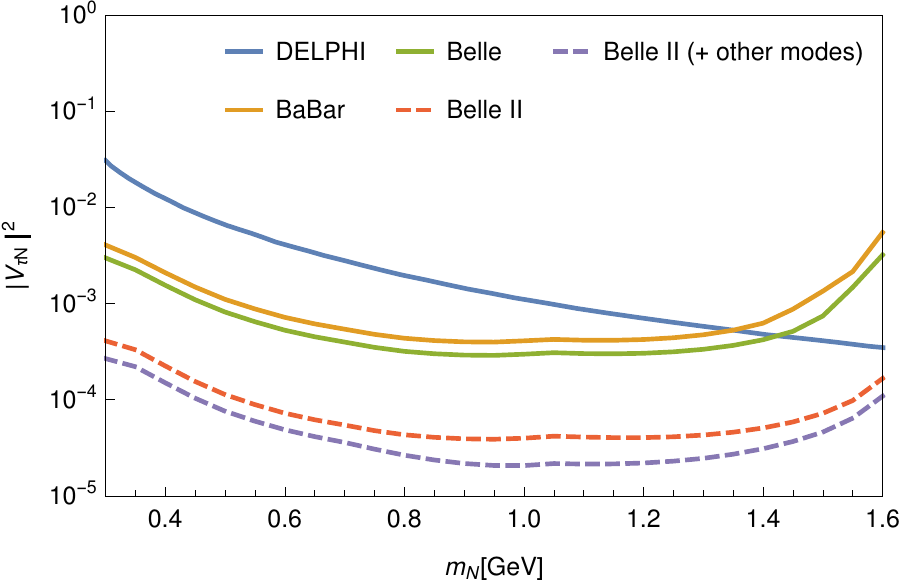}
\caption{Expected 95\% confidence-level limits on the coupling $|V_{\tau N}|^2$ vs. the $N$ mass, obtainable from $\tau^-\to X_1 N$, $N\to X_2 \nu$ for $X_1=\pi^-(\pi^0)$ and $X_2=\ell^+\ell^-$ at BaBar (yellow), Belle (green) and Belle~II (red).
Also shown is the potential impact of adding the modes $X_1 = \ell\nu$ and $X_2 = \pi^+ \pi^-$.
For comparison we also plot current limits from DELPHI \cite{Abreu:1996pa}.
}
\label{fig:limitsv2}  
\end{figure}

Conducting these fits is beyond the scope of this study. Rather, we estimate the results  of a simple cut in lieu of a fit. Using the distributions of Fig.~\ref{fig:distr} we find that a bin-by-bin cut that retains 90\% of the signal events  rejects over 97\% of the background. The combination of these simple cuts on $E_2$ vs. $E_1$ and $m_2$ vs. $m_1$ retains 6 background events in the entire Belle~II sample. From this, we conclude that a more sophisticated analysis with the full Belle~II data set can come close to being background-free, and  assume an additional efficiency loss of 75\% to account for these cuts. Based on this  rough estimate, we show in Fig.~\ref{fig:limitsv2} the expected limits on $|V_{\tau N}|^2$ as a function of mass, given the data samples of the  BaBar, Belle, and Belle~II experiments. Fig.~\ref{fig:limitsv2} also shows the impact of using the final states decays $X_1=\ell\nu$ and $X_2=\pi^+\pi^-$, assuming the same signal efficiency as for the $X_1=\pi$, $X_2=\ell^+\ell^-$ mode. The branching fraction for $N\to \nu \pi^+\pi^-$ is taken from Ref.~\cite{1805.08567}. 

In summary, we propose a new search for a sterile neutrino $N$ 
with $m_N < m_\tau$
that mixes predominantly with the $\tau$ neutrino. 
Having negligible mixing with the $\nu_e$ and $\nu_\mu$, $N$ evades detection in most searches that require a final-state lepton. The current best limits, obtained by DELPHI, can be surpassed by carrying out the proposed search at current and near-future $B$-factories, making use of their large samples of $e^+e^-\to \tau^+\tau^-$ events to produce the $N$ via  $\tau\to X_1 N$ decays. 
Our method exploits the long lifetime of $N$ in this mass range to greatly suppress background. We propose kinematic and vertex-based constraints to further suppress background and measure the $N$ mass if signal is observed. Since these constraints hold when $X_1$ a hadronic system, we do not study the case $X_1=\ell\nu$ in detail. Since this case roughly doubles the available branching fraction and does not suffer from the $\tau\to \pi K_L \nu$ background, we include it in the final sensitivity estimation with the caveat that it requires further study.

This work was supported by FONDECYT (Chile) grants 1170171, 3170906, 1161463 and 3180032, and CONICYT (Chile) PIA/Basal FB0821, by grants from ISF (2181/15, 2476/17), GIF (I-67-303.7-2015), BSF (2016113) (Israel), and by the European Union’s Horizon 2020 research and innovation programme under the Marie Skłodowska-Curie grant agreement No. 822070.

\end{document}